\documentclass[a4paper, reqno, final]{dmtcs-episciences}

\usepackage[latin1]{inputenc}
\usepackage[T1]{fontenc}
\usepackage{lmodern}
\usepackage[draft=false,kerning=true]{microtype}
\usepackage{subcaption}

\usepackage{a4wide}

\usepackage{amsmath, amssymb, amsthm, amsfonts, amscd}

\usepackage{graphicx,color}
\usepackage{tikz}
\usetikzlibrary{automata}

\usepackage{fancyvrb}

\usepackage{ifpdf}
\ifpdf
\fi

\theoremstyle{definition}

\newcommand{\SageMath}{\textsf{SageMath}}
\newcommand{\SageMathCloud}{\textsf{SageMathCloud}}
\newcommand{\SageTeX}{\textsf{Sage}\TeX}

\newcommand{\TikZ}{\textsf{Ti\textit{k}Z}}

\RequirePackage{bm}
\bmdefine{\bfldot}{.}
\renewcommand{\centerdot}{\bfldot}

\newcommand{\MR}[1]{}

\newcommand{\skipnb}{}
\newcommand{\nbmd}[1]{}
\newcommand{\nbcode}[1]{}

\newcommand{\TODO}[1]%
{\par\fbox{\begin{minipage}{0.9\linewidth}\textbf{TODO:} #1\end{minipage}}\par}

\skipnb\title[Automata in SageMath]{Automata in SageMath---Combinatorics meets \\
  Theoretical Computer Science}

\author[Clemens Heuberger, Daniel Krenn, Sara Kropf]{
  Clemens Heuberger\affiliationmark{1} \and
  Daniel Krenn\affiliationmark{1} \and
  Sara Kropf\,\affiliationmark{1}%
  \footnotetext{The authors are supported by the Austrian Science Fund (FWF):
    P~24644-N26 and by the Karl Popper Kolleg
    ``Modeling--Simulation--Optimization''
    funded by the Alpen-Adria-Universität Klagenfurt and
    by the Carinthian Economic Promotion Fund (KWF).}%
  \footnotetext{Email-addresses:
    \href{mailto:clemens.heuberger@aau.at}{clemens.heuberger@aau.at},
    \href{mailto:math@danielkrenn.at}{math@danielkrenn.at} \textit{or}
    \href{mailto:daniel.krenn@aau.at}{daniel.krenn@aau.at},
    \href{mailto:sara.kropf@aau.at}{sara.kropf@aau.at}}}

\affiliation{Institut f\"ur Mathematik,
  Alpen-Adria-Universit\"at Klagenfurt
}


\usepackage{sagetex}
\makeatletter
\renewcommand{\sagestr}[1]{\ST@sage{str(#1)}}
\makeatother

\makeatletter
\def\sageverb
{%
  \relax\leavevmode\null
  \bgroup
  \verbatim@font
  \let\do\do@noligs  \verbatim@nolig@list
  \let\do\@makeother \dospecials
  \@vobeyspaces
  \@sageverb
}%
\def\@sageverb #1{\catcode`#1\active
                \lccode`\~`#1%
                \lowercase{\long\def~##1~}{\sagestr{##1}\egroup}%
                \lowercase{~}}%
\makeatother

\newenvironment{sageverbenv}{%
\medskip\par\noindent\hspace{\sagetexindent}\begin{minipage}{\textwidth}}%
{\end{minipage}\medskip}  

\newenvironment{riddle}{\begin{quote}\itshape}{\end{quote}}

\begin{document}
\VerbatimFootnotes


\received{2016-01-13}

\revised{2016-04-13}

\accepted{2016-05-02}

\publicationdetails{18}{2016}{3}{9}{1352}


\maketitle

\begin{abstract}
  The new finite state machine package in the
  mathematics software system \textsf{SageMath} is presented and illustrated by
  many examples. Several combinatorial problems, in particular digit
  problems, are introduced,  modeled by automata and transducers and solved using 
  \textsf{SageMath}.

  In particular, we compute the asymptotic
  Hamming weight of a non-adjacent-form-like digit expansion, which was not
  known before.
\end{abstract}  


\nbmd{<h1>Automata in SageMath---Combinatorics meets Theoretical Computer Science</h1>
<i>Clemens Heuberger, Daniel Krenn, Sara Kropf</i>}
\nbmd{Note. To run this worksheet you need to be logged in and to make
  a copy of the worksheet (otherwise, it is read only). You need at
  least SageMath 6.10.}


\bgroup\renewcommand{\figurename}{Figure}

\setcounter{section}{-2}
\let\thesectionnormal\thesection
\def\thesection{0}
\skipnb\section{Prelude}
\label{sec:intro}


\skipnb\subsection{First Act: Digits and Multiplication}


Ladies and Gentlemen, let me introduce you to the first guest of this
opening ceremony: the equation
\begin{equation*}
  9000 + 900 + 90 + 9 = 10000 - 1.
\end{equation*}
You might ask now: Why?---Why exactly this?---A small hint:
It has to do with multiplication as the title of the first act
says. Already guessed?---Yes?---No?---Then, here is the answer.

Suppose that we want to multiply a huge number~$n$ by $9999$. Suddenly today's
guest list becomes clearer. Such a multiplication should not be
done according to the schoolbook, but just by subtracting $n$ from
$10000n$. Since nowadays such calculations are done by computers, we
switch to binary digit expansions at this point. That is our cue to 
introduce the next guests: the digits $0$ and
$1$. Usually they are used when writing a number in standard binary expansion,
for example
\begin{equation*}
  14 = 1\cdot 8 + 1\cdot 4 + 1\cdot 2 + 0\cdot 1
  = (1110)_2.
\end{equation*}
Tonight they will perform in a special way, but they need help from their digit
sibling $\bar{1}=-1$, which brings us to the following riddle:
\begin{riddle}
  Write $14$ in base~$2$ with the digits $-1$, $0$ and $1$  
  such that there are fewer nonzero digits than in the standard binary
  expansion.
\end{riddle}
Our digit triplet needs only a moment to prepare a solution.
The lights go on again and we immediately see
\begin{equation}\label{eq:NAF12-intro}
  14 = 1\cdot16 + 0\cdot8 + 0\cdot4 - 1\cdot2 + 0\cdot1
  = (100\bar{1}0)_2.
\end{equation}
After thinking about this for a moment, we deduce that it cannot work with only
one nonzero, so we are satisfied with this perfect solution. This brings us
directly to the special guest of this evening, the non-adjacent form,
which will be present in the whole article. It guarantees that no two
adjacent digits are nonzero at the same time; and the digit
expansion~\eqref{eq:NAF12-intro} is indeed such a non-adjacent form.
A bunch of questions appear:
\begin{riddle}
  So how do we determine such a digit expansion of a general number?
  Is there a way to simply get it from the binary expansion? Is this
  the best possible way to write an integer?
\end{riddle}
Questions over questions---all of which will be answered after a short
break.


\skipnb\subsection{Second Act: Automata and Other Boxes}


In this second act we will first meet the main characters of this
article. The multiplication, expansions and digits from the previous
act still play their role, which is to be nice examples. As a small
side note, the authors Clemens, Daniel and Sara work with digit
expansions a lot in their research, so these examples seem to be a
natural choice.

Back to business; there are some presents for each of you---these
little boxes over there called automata. You will find them everywhere. Some of
these magic boxes can only be opened giving them a special digit
expansion (see Section~\ref{sec:recogn-naf}). And, it becomes better and better. Some of them
even speak some words; they are known as transducers and will return,
for example, a non-adjacent form for its binary input.  The trick
behind these magic boxes will be revealed in
Section~\ref{sec:what-an-automaton}; even better you can take a look
inside them, even see how they are constructed and create new such
boxes in Sections~\ref{sec:non-adj-form}, \ref{sec:3-transducers} and~\ref{sec:32-12-NAF}.


\skipnb\subsection{Third Act: Strange New World}


Welcome back everybody to this third act. The next
guest was already present, but hidden behind the scenes. It offers an
immense amount of mathematical objects combined with algorithms for
working with them. All the automata and transducers mentioned
previously are integrated in it. Contributions to its code-base are
subjected to a transparent peer-review process. It stands for reproducible
results as all parts of its library are tested with each future release.
For the last couple of
years, more and more researchers are using it. Ladies and Gentlemen,
we proudly present the free and open-source mathematics software system
\SageMath{}~\cite{SageMath:2015:6.10}.

There are great news these days: \SageMath{} has now a module for
finite state machines, automata, and transducers~\cite{Heuberger-Krenn-Kropf:2013:fsm-sage},
which was written and is still improved by the three authors of this article.
One of the main motivations of using \SageMath{} as the home
for this finite state machine module was to allow the use of
more or less everything in the zoo of 
mathematical objects available in \SageMath{}, in particular during
the construction and
manipulation of automata and transducers. In general, the use
of tools from different areas is one of the big strengths of \SageMath{}.

The aim of this piece of work is to demonstrate the functionality of
the new finite state machines package in the form of a tutorial.
The interplay between automata and transducers on the one hand and,
for example, graph theory, linear algebra, and asymptotic analysis on
the other hand will be delineated and worked out. A more detailed
overview can be found in Section~\ref{sec:in-this-tutorial}.


\skipnb\subsection{Fourth Act: What Next?}


Now, in this final act of the prelude, we make things more concrete.
Let us have another look at the binary expansion
and the non-adjacent form of integers. Intuitively it is clear that on
average one half of the digits in a binary expansion are $1$ and,
because of the restrictions in non-adjacent forms, these expansions
should have less nonzeros. But what is the true value?---Can someone
guess this number?---Maybe one third of the digits are nonzeros?---Yes,
indeed, this is correct, however, it is not a consequence of having used
exactly $3$ digits, but a coincidence.
So, guess what does it take to calculate these
numbers and their corresponding asymptotic expansions?

Typically (weighted) adjacency matrices of the underlying graphs of a
transducer and their eigenvalues are used, and indeed \SageMath{} can
do this out of the box (see the finale on the last couple of pages
 of this article). Another example is testing whether the
non-adjacent form of any integer minimizes the number of nonzero
digits. Huh!?---How to do this?---The answer is: Construct a suitable transducer,
use the underlying digraph and apply the Bellman--Ford algorithm on it,
\dots{} Eureka!

We are already in the final phase of this introduction and there is
something missing. So far all these quantities and results were known
before. But what about finding new results to open
problems?  This brings us to our surprise guest, which is related to
the non-adjacent form, but uses digits $-2$ and $2$ as well and is
constructed in a special way. Is this a good expansion to use for our
multiplication problem mentioned in the first act? How do we compute
it, and is it the best possible? What about its average number of
nonzeros? Many questions here, but all will be answered
illustratively in Section~\ref{sec:32-12-NAF}.

There are only a few things left to say. Thank you for reading these
first four acts and stay with us to get to know the really cool, interesting
and practical stuff. Enjoy the rest of the evening!


\def\thesection{$\frac12$}
\skipnb\section{Bonus: Questions to and Answers from the Authors}


\skipnb\subsection{How Do I Get This Awesome New Finite State Machines Package?}


To keep it short: It is already included in \SageMath{}.\footnote{The basic version
  was included in \SageMath{}~5.13~\cite{Stein-others:2013:sage-mathem-5.13}
  (for details see the relevant
  ticket~\cite{Heuberger-Krenn-Kropf:2013:fsm-sage} on the \SageMath{} trac
  server). To work with
  all features of this tutorial, use \SageMath{}~6.10 or later.}
If you are using the finite state machines
package of \SageMath{} in your own work, please let us know.


\skipnb\subsection{Can I Easily Use SageMath?}

Definitely yes, it is very simple to use \SageMath. All you need is a
modern web-browser, going to
\url{https://cloud.sagemath.com}, create a free
account and start playing around in the \SageMathCloud.
Even better: All the code examples of this article are available
there.\footnote{A worksheet with all code examples is available in the
  \SageMathCloud{} at
  \url{https://cloud.sagemath.com/projects/9a0d876c-4515-4a30-968a-96bd139c3c19/files/fsm-in-sage/fsm-in-sage.sagews}}

Of course and if you prefer, \SageMath{} can be run locally on
your computer as well.


\skipnb\subsection{What is This Tutorial about?}
\label{sec:in-this-tutorial}


As mentioned in the prelude,
the aim of this piece of work is to demonstrate some of the functionality of
the new finite state machines package in the form of a tutorial. Detailed
documentation of the available methods and their parameters can, as usual, be
found in the \SageMath{} documentation.\footnote{The \SageMath{} documentation
  of the finite state machine module can be found at
  \url{http://doc.sagemath.org/html/en/reference/combinat/sage/combinat/finite_state_machine.html}.}
There, further examples are presented as well.

Along the way, a new result on a digit system related to the non-adjacent
form~\cite{Reitwiesner:1960} is proved.
We consider a new digit expansion with base~$2$ and digit set $\{-2,-1,0,1,2\}$.
We compute the expected
value of the Hamming weight, i.e., the number of nonzero digits, of this digit
expansion of integers less than $2^{k}$. Although the digit set is larger
than the one for the non-adjacent form, it turns out that the expected value of
this new digit expansion is worse than that of the standard binary expansion
and therefore worse than that of the non-adjacent form.


\skipnb\subsection{Where Do I Find What?}
\label{sec:where-what}

While answering this question, we give a brief overview on how to show the
result mentioned above. We also demonstrate the enormous advantages of using
\SageMath{} for constructing and simulating automata and transducers.
We start at the
beginning and get to know automata in
Section~\ref{sec:what-an-automaton}. We will see various ways to
construct them and use them as building blocks to construct larger and better
machines (the whole Section~\ref{sec:non-adj-form}). Some first
insights in the interplay between the concept of automata and
combinatorics are found in Section~\ref{sec:counting}.

We then go on to the
generation of transducers, which can be done in several ways as well.
First, we can simply list all
transitions of a finite state machine (cf.\
Section~\ref{sec:non-adj-form-manually}). Second, we can use transition
functions written in \SageMath{} to construct a transducer (cf.\
Section~\ref{sec:NAF-transition-function}, but we will use it in several other
places as well). This is of course possible for any type of finite state
machine. Another way is to construct machines by a suitable
combination of smaller
building blocks: We manipulate and combine several finite state machines
properly (Sections~\ref{sec:NAF-minus-construction}
and~\ref{sec:construction-32-12-NAF}). A couple of variants and more advanced
constructions will also be shown (see, for example,
Sections~\ref{sec:what-makes-naf-better} and~\ref{sec:recognizing}).

Equipped with the wisdom gained up to this point, we go on to a new
challenge. Section~\ref{sec:32-12-NAF} is devoted to a new digit expansion
related to the non-adjacent form. We want to analyze it and investigate its properties. The main result will be the asymptotic behavior
of the average output (cf.\ Section~\ref{sec:weight-more}).
This can be obtained by just one function call. However, behind the scenes, the
interplay between automata and transducers on the one side and other areas
of mathematics within \SageMath{} on the other side plays an essential role.
Linear algebra with 
adjacency matrices (Section~\ref{sec:adj-matrix}) and asymptotic analysis
will appear.

One final remark before getting started: All of these steps can be
computed within \SageMath{}. Thus, we can use everything offered by
the mathematical software system and construct powerful finite state
machines for the analysis, without shuffling data from one system to
another.


\let\thesection\thesectionnormal

\section{Using an Automaton Properly}
\label{sec:what-an-automaton}

You are already well acquainted with automata and transducers?---Great, skip to
Section~\ref{sec:non-adj-form}. Otherwise, stay with us, a gentle introduction
to these useful machines is provided here.

Figure~\ref{fig:ex-automat} shows an example of an \emph{automaton}. Obviously,
an automaton has little to do with a vending machine which returns a snack
after inserting a coin. So what shall we stuff in instead and what bubbles out?

This is easy to explain: The input is a sequence of numbers, letters, etc.
The automaton then returns an answer ``Yes, I accept this input,'' or
``No, I do not accept this input.'' So apparently there is something going on,
and we want to understand:
How does the automaton know which answer to give?

Let us say we insert the input sequence $0,-1,0,0,1$ in the automaton of
Figure~\ref{fig:ex-automat}. There is one arrow entering this picture;
this is at the \emph{initial state}~$0$ where we start. The
automaton reads the first element $0$ of the input
sequence and then decides to which state it goes next. It chooses
the \emph{transition} (i.e.\ a labeled arc) starting in
the \emph{current state}~$0$ whose label equals the current input~$0$.
This loop leads to state~$0$ again, which is now our new current state.
We proceed as above and read the next
element~$-1$ of the sequence. This leads to state~$1$.
The next elements are~$0$, which leads back to state~$0$,
another~$0$, which keeps us in state~$0$, and
finally $1$, which takes us to state~$1$.

\begin{figure}
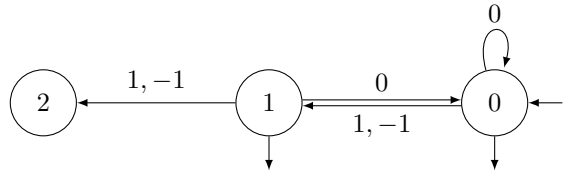

  \centering
  \begin{sagesilent}
    # Figure: An example of an automaton
    w = automata.Word
    NAF_language = ((w([0]) + w([1, 0]) + w([-1, 0])).kleene_star() *
                    (automata.EmptyWord([-1, 0, 1]) + w([1]) + w([-1]))
                   ).minimization().coaccessible_components().relabeled()
    NAF_language.add_state(2)
    NAF_language.add_transition(1, 2, 1)
    NAF_language.add_transition(1, 2, -1)
    NAF_language.latex_options(
        initial_where=lambda _: 'right',
        coordinates={0: (6, 0), 1: (3, 0), 2: (0, 0)},
        accepting_style='accepting by arrow',
        accepting_where={0: 'below', 1: 'below'})
  \end{sagesilent}
  \sagestr{latex(NAF_language)}
  \caption{An example of an automaton.}
  \label{fig:ex-automat}
\end{figure}


As there is no more input, the automaton will stop and return an answer:
State~$1$ is a \emph{final state} (in Figure~\ref{fig:ex-automat}
marked by a small outgoing arrow), thus the answer is
``Yes, I accept this input.''

If the last current state is not final, then the answer is
``No''. Moreover, if there is no valid transition to choose, then the
answer is ``No'' as well. These cases are illustrated by the
following examples. Suppose we use the input sequence $1,1$ and
the automaton in Figure~\ref{fig:ex-automat}. The negative answer is returned
since we end up in state~$2$, which is nonfinal.
Using the input sequence $1,1,0$, the
answer is still ``No'', but for a different reason. State~$2$ is reached
by the first two~$1$, and then a~$0$ is read. There is no transition
starting in state~$2$ with label~$0$, so the automaton cannot
proceed further.

So much for an informal description.\footnote{Of course, it is possible to
  define an automaton formally as a quintuple consisting
of a set of states, a set of initial states, a set of final
states, an alphabet and a transition function (see Hopcroft, Motwani and Ullman~\cite{Hopcroft-Motwani-Ullman:2001:automata}
or Sakarovitch~\cite{Sakarovitch:2009:elemen}). But for this tutorial,
it is enough to think of an automaton as a graph with additional
labels as in Figure~\ref{fig:ex-automat}.}
You may have got the impression that automata are not very entertaining
because they can only answer ``Yes'' or ``No''. If so, we want to
change this---there exist transducers as well, which
provide a more sophisticated output.
In principle, a \emph{transducer} is the same as an automaton with
an additional second label (the \emph{output label}) for every
transition. Figure~\ref{fig:ex-transducer} shows such an animal: For
example, the label
$1\mid 0$ means that this transition has an input label~$1$ (as used
by automata when reading the input sequence) and an output label~$0$.
Every time this transition is used, the output~$0$ is
written at the end of the previous outputs. There is one new symbol, namely
an output label~$\varepsilon$, which stands
for the empty word (i.e., nothing is written).
Let us illustrate this by an example, too.
For the transducer in Figure~\ref{fig:ex-transducer}, the input sequence
$0,1,1,1,0,0$ produces ``Yes'' together with the output
sequence~$0,-1,0,0,1$ (and note that our first output was an $\varepsilon$).

\begin{figure}
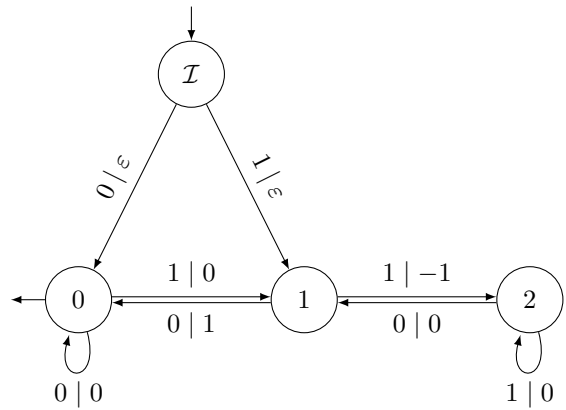

\centering
\begin{sagesilent}
  # Figure: An example of a transducer
  NAF0 = Transducer([(-1, 0, 0, None), (-1, 1, 1, None), (0, 0, 0, 0),
                     (0, 1, 1, 0), (1, 0, 0, 1), (1, 2, 1, -1),
                     (2, 1, 0, 0), (2, 2, 1, 0)],
                     initial_states=[-1], final_states=[0],
                     input_alphabet=[0, 1])
  NAF0.state(-1).format_label=lambda: r'\mathcal{I}'
  NAF0.latex_options(
      coordinates={-1: (1.5, 3),
                   0: (0, 0),
                   1: (3, 0),
                   2: (6, 0)},
      initial_where={-1: 'above'},
      loop_where=lambda x: 'below',
      accepting_style='accepting by arrow',
      accepting_where={0: 'left'})
\end{sagesilent}
\sagestr{latex(NAF0)}
\caption{An example of a transducer.}\label{fig:ex-transducer}
\end{figure}

It is almost time to get our hands dirty, so let us pose a couple of
questions: How do we get a transducer like the one
in Figure~\ref{fig:ex-transducer}?  Do we have to construct it with pen and
paper? Luckily, the answer is no; \SageMath{} can do this construction
for us automatically. We will see this in the next sections.


\section{Making Us Comfortable with the Non-Adjacent Form as a Warm-Up}
\label{sec:non-adj-form}


Before we start constructing automata---and later on transducers---let us
have a look at the terms used in this
section. We start by explaining the classical \emph{non-adjacent form}  of an
integer,
abbreviated as \emph{NAF}. It is a representation with base~$2$ and digits $-1$, $0$ and $+1$,
such that two adjacent digits are not both nonzero.
By setting $\bar{1}=-1$, this means that we forbid all blocks
$11$, $\bar{1}1$, $1\bar{1}$ and $\bar{1}\bar{1}$ in the digit expansion. For example, we have
\begin{equation}\label{eq:NAF14}
  14 = 1\cdot16 + 0\cdot8 + 0\cdot4 - 1\cdot2 + 0\cdot1  = (100\bar{1}0)_2.
\end{equation}
In contrast, $14 = (10\bar{1}10)_2$ is not a non-adjacent form.
It can easily be shown that each integer has a unique representation by a NAF,
cf.\ Reitwiesner~\cite{Reitwiesner:1960}.

Usually, digit expansions are written from the most significant digit (on the
left-hand side) to the
least significant digit (on the right-hand side).
As we compute digit expansions exactly in the other direction in this article\footnote{This
  approach uses information modulo the base. An alternative approach,
  not adopted here, uses greedy expansions, where digits are computed from the
most significant to the least significant one. }, we also write them
in that other direction when used as input or output of finite state machines.  Therefore, we have two different notations. For digit expansions, like in
$(100\bar{1}0)_2$, we write the least significant digit on the
right. For inputs and outputs of finite state machines, like in 
\texttt{[0, -1, 0, 0, 1]}, we write the least significant digit on the
left. Consequently, we also speak of trailing zeros if we append zeros after
the most significant digit.


\subsection{Accepting Everything}
\label{sec:recogn-naf}

It is time to create our first automaton. We have already learned
about the non-adjacent form, so let us construct an automaton recognizing
such digit expansions. So a better title of this section would have been ``Accepting
Something''.

The language of non-adjacent forms (with possible trailing zeros) can be written as the regular expression
\begin{equation*}
  (0+10+\bar10)^{*}\,(1+\bar1+\varepsilon)
\end{equation*}
meaning: A non-adjacent
form starts with $0$, $10$ or $\bar10$, where ``or'' is usually
denoted by $+$. This part can
be repeated arbitrarily often (also zero times), which is denoted by
the Kleene star
${}^{*}$. Up to now, we get a valid non-adjacent form since each
nonzero is followed by a zero, but there are some expansions
missing. So at the end we might have to add a $1$ or $\bar1$, or not
(denoted by the empty word $\varepsilon$).

Such regular expressions are closely related to automata; even better, there
is a direct way to translate them to obtain an automaton which accepts
non-adjacent forms. We type\footnote{This document was created using \SageTeX,
  which comes along with \SageMath{}. It allows the following: We type \SageMath{}
  source code directly in the \TeX-document, this code is then executed by
  \SageMath{} and the corresponding outputs (results) are typeset here.}
\begin{sageblock}
  one = automata.Word([1])
  zero = automata.Word([0])
  minus_one = automata.Word([-1])
  epsilon = automata.EmptyWord(input_alphabet=[-1, 0, 1])
  NAF_language = ((zero + one*zero + minus_one*zero).kleene_star() *
                  (one + minus_one + epsilon)).minimization()
\end{sageblock}
In fact, we have now created many (smaller) automata\footnote{For example,
  \verb|one = automata.Word([1])| is an automaton which only accepts the
  input word of the single letter~$1$ and nothing else.
  A \verb|*| concatenates two automata.
  The method \texttt{minimization()} returns a minimal
  version of the given automaton. If you want to know more about how
  it works, consult the documentation of \SageMath, for example, by
  typing \verb|NAF_language.minimization?|\,.}
and have joined them to get the automaton \verb|NAF_language|.

For now the result is stored in the variable \verb|NAF_language|. By
typing \verb|NAF_language| we get
\begin{sageverbenv}
  \verb+  +\sageverb|NAF_language|
\end{sageverbenv}
Since a picture says a thousand words,
let us draw this automaton\footnote{More precisely,
  Figure~\ref{fig:naf-automaton-complete} shows
  \verb|NAF_language.relabeled()|; this
  provides nice labels.}
in Figure~\ref{fig:naf-automaton-complete}. Hey!---State~$2$ of this automaton is a black
hole, a so-called sink: Once an input sequence reaches this state, it is lost forever. Can't we get rid
of such states?---Of course; the command is
\begin{sageblock}
  NAF_language_coaccessible = NAF_language.coaccessible_components()
\end{sageblock}
and the result is shown in Figure~\ref{fig:naf-automaton-coaccessible}.
 More on drawing finite state machines
comes later in Section~\ref{sec:plot}.

\begin{figure}
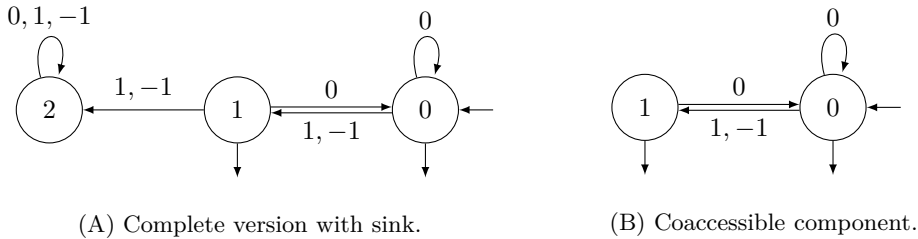

  \centering
  \begin{sagesilent}
    # Figure: The automaton NAF_language which accepts
    #         non-adjacent forms
    NAF_language_picture = NAF_language.relabeled()
    NAF_language_picture.latex_options(
        coordinates={0: (0, 0), 1: (-2.5, 0), 2: (-5, 0)},
        initial_where={0: 'right'},
        accepting_style='accepting by arrow',
        accepting_where={0: 'below', 1: 'below'})
    # Figure: The automaton NAF_language which accepts
    #         non-adjacent forms; only coaccessible states
    NAF_language_coaccessible_picture = NAF_language_coaccessible.relabeled()
    NAF_language_coaccessible_picture.latex_options(
        coordinates={0: (0, 0), 1: (-2.5, 0)},
        initial_where={0: 'right'},
        accepting_style='accepting by arrow',
        accepting_where={0: 'below', 1: 'below'})
  \end{sagesilent}
  \begin{subfigure}{0.45\linewidth}
  \centering
  \sagestr{latex(NAF_language_picture)}
  \caption{Complete version with sink.}\label{fig:naf-automaton-complete}
  \end{subfigure}
  \begin{subfigure}{0.45\linewidth}
  \centering
  \sagestr{latex(NAF_language_coaccessible_picture)}
  \caption{Coaccessible component.}\label{fig:naf-automaton-coaccessible}
  \end{subfigure}
  \caption{The automaton \texttt{NAF\_language} which accepts non-adjacent forms.}
  \label{fig:naf-automaton}
\end{figure}

Maybe this is a good point to mention that a couple of
commonly used automata and transducers, like \verb|automata.Word|, are already prebuilt in \SageMath{}
and can be used directly; take a
look into the \SageMath{} documentation.

So, we have our first own creation of an automaton.---Now we want to see it
working. The command
\begin{sageverbenv}
  \verb+  +\verb|NAF_language([0, -1, 0, 0, 1])|
\end{sageverbenv}
\begin{sagesilent}
  expansion_14_NAF = [0, -1, 0, 0, 1]
  expansion_14_bin = [0, 1, 1, 1]
\end{sagesilent}
returns \sageverb|NAF_language(expansion_14_NAF)|, i.e., it accepts the input
as it should. In contrast,
\begin{sageverbenv}
  \verb+  +\verb|NAF_language([0, 1, 1, 1])|
\end{sageverbenv}
returns \sageverb|NAF_language(expansion_14_bin)|, so the input is
rejected by the automaton. We lean back and are
satisfied for now.


\subsection{Accepting Everything a Second Time}
\label{sec:recogn-naf-complement}


Our successful first construction motivates us to proceed. Letting our minds
flow, we remember that non-adjacent forms are also characterized by
forbidding each of the subwords $11$, $\bar11$, $1\bar1$ and
$\bar1\bar1$. Shouldn't this fact be used somehow?

Looking up some prebuilt automata and ways to transform them, we think that
\begin{sageblock}
  automata.ContainsWord([1, 1], input_alphabet=[-1, 0, 1]).complement()
\end{sageblock}
will be helpful. This machine accepts words without the subword $1,1$.
As a small explanation, \verb|complement()|
constructs an automaton which accepts exactly the input sequences
which were rejected by the previous automaton. It seems to be logical
to copy and paste this command to create automata dealing with the other forbidden subwords. But what now?---We need some intersection of them.---No problem at all, just use the operator \verb|&| for intersection!

So all together we type
\begin{sageblock}
  NAF_language2 = \
    (automata.ContainsWord([1, 1],   input_alphabet=[-1, 0, 1]).complement() &
     automata.ContainsWord([1, -1],  input_alphabet=[-1, 0, 1]).complement() &
     automata.ContainsWord([-1, 1],  input_alphabet=[-1, 0, 1]).complement() &
     automata.ContainsWord([-1, -1], input_alphabet=[-1, 0, 1]).complement()
    ).minimization()
\end{sageblock}
Is this the same as our first automaton \verb|NAF_language|? The name and our
ideas for the construction suggest
that it is. So let use check this by
\begin{sageblock}
  NAF_language2.is_equivalent(NAF_language)
\end{sageblock}
This gives---suspense builds---the result
\skipnb\sageverb|NAF_language2.is_equivalent(NAF_language)|. Puh, what a relief!

Equivalently, by De Morgan's law, we ``negate'' this construction and
forbid all words which contain any of
the subwords $11$, $\bar11$, $1\bar1$, or $\bar1\bar1$. This is done via
\begin{sageblock}
  NAF_language3 = \
    (automata.ContainsWord([1, 1],   input_alphabet=[-1, 0, 1]) +
     automata.ContainsWord([1, -1],  input_alphabet=[-1, 0, 1]) +
     automata.ContainsWord([-1, 1],  input_alphabet=[-1, 0, 1]) +
     automata.ContainsWord([-1, -1], input_alphabet=[-1, 0, 1])
    ).determinisation().complement().minimization()
\end{sageblock}
Again \verb|NAF_language3.is_equivalent(NAF_language)| yields
\sageverb|NAF_language3.is_equivalent(NAF_language)|, so we have done
everything correctly.

But wait!---What exactly did we do?---What is \verb|determinisation()|? And
why do we need it? The answer to the second question is that the
result of the operation \verb|+| is not
deterministic.\footnote{A nondeterministic automaton can have more than
  one possible path for an input sequence. An input is accepted
  if there is at least one path with this input
  sequence as labels leading to a final state.} And
\verb|determinisation()| is an algorithm which makes this automaton
deterministic. Some fact: For each automaton an equivalent
deterministic automaton can be created. But be aware that the
determinisation of an automaton can increase the number of states
exponentially. That is the reason why it is not done automatically,
and you have to call it yourself.


\subsection{It's Counting Time}
\label{sec:counting}


As interesting as the previous tasks were, we want more mathematics
and make a small excursion
now: It is counting time! We ask: How many NAFs of a given length are there?

Once we have an automaton accepting the NAF, \SageMath{} should be able tell us the answer. And indeed, this can be achieved by typing
\begin{sageblock}
  var('n')
  N = NAF_language.number_of_words(n)
\end{sageblock}
The answer is
\begin{equation*}
  \sage{N},
\end{equation*}
which seems like magic.---How does this trick work? This will be revealed
in Section~\ref{sec:adj-matrix}.


This is the end of our section on automata. We come back to automata a
bit later in Section~\ref{sec:accept-naf-from-transducer} because there is a further possibility creating a NAF-automaton:
We use the output
of a transducer which transforms any digit expansion with digits $-1$,
$0$ and $1$ into a non-adjacent form and make an automaton out of it.

So, what comes next?---We already have methods that tell us whether
some input is a non-adjacent form and how many there are, but we also
want to calculate these digit expansions. This brings us to the
transducers of the following sections.


\section{Three Kinds of Calculating the Non-Adjacent Form}
\label{sec:3-transducers}


We already heard an introduction to non-adjacent forms at the beginning
of Section~\ref{sec:non-adj-form}, and building
various automata accepting these digit expansions went quite well. So let us now
tackle the question: How do I calculate a non-adjacent form?


\subsection{Creating a Transducer from Scratch}
\label{sec:non-adj-form-manually}


In~\cite[Figure~2]{Heuberger-Prodinger:2006:analy-alter}, a transducer for
converting the binary expansion of an integer $n$ into its non-adjacent form is
given. We reproduce it here as Figure~\ref{fig:ex-transducer} and directly translate it
into \SageMath. We write
\begin{sageblock}
  NAF1 = Transducer([('I', 0, 0, None), ('I', 1, 1, None),
                     (0, 0, 0, 0), (0, 1, 1, 0),
                     (1, 0, 0, 1), (1, 2, 1, -1),
                     (2, 1, 0, 0), (2, 2, 1, 0)], 
                    initial_states=['I'], final_states=[0], 
                    input_alphabet=[0, 1])
\end{sageblock}
to construct this transducer with states \verb|'I'| (the string \texttt{I}),
\verb|0|, \verb|1| and \verb|2| (the integers $0$, $1$ and $2$, respectively).
The list of $4$-tuples defines the transitions of the transducer. For example,
\verb|(1, 2, 1, -1)| is a transition from state~$1$ to state~$2$ with
input~$1$ and output~$-1$. Here, input means reading a (more
precisely, the next) digit of the binary expansion of~$n$. The output is a
digit of the non-adjacent form, written step-by-step. Note that we read and
write the expansions from the least significant digit to the most significant
one and we start at the digit corresponding to $2^0=1$.

This approach required us to manually model the digit conversion as a
transducer (or, in this particular instance, to find a reference in the
available literature). Shouldn't
there be an easier method using the full power of \SageMath{}? For sure
there is.---And we will
do so in later examples to demonstrate various approaches for constructing
transducers.


\subsection{The Non-Adjacent Form of Fourteen}
\label{sec:desc-NAF}


\begin{sagesilent}
  NAF = NAF1
\end{sagesilent}
Let us
compute the non-adjacent form of, for example, our lucky number
fourteen.\footnote{You may ask why $14$ is our lucky number. In fact, it is not!
  But it is not that bad. This number is, beside the actual $13$, our second
  lucky number. The reason of preferring $14$ over $13$ is simple and of
  educational character: The digit expansion (used in this tutorial) of $13$ is
  too symmetric. This may lead to confusion whether this expansion is read
  from left to right or the other way round.}
We intend to use the transducer above, so,
for convenience, we set \verb|NAF = NAF1|.
Remember that you can see the transducer \verb|NAF| in
Figure~\ref{fig:ex-transducer}.

As a first try, we type\nbcode{NAF_of_14 = NAF(14.digits(base=2))}
\begin{sageverbenv}
  \verb+  +\verb|NAF_of_14 = NAF(14.digits(base=2))|
\end{sageverbenv}
and get
\begin{sageverbenv}
  \verb+  +\verb|ValueError: Invalid input sequence.|
\end{sageverbenv}
An error message?---Huh?---So did we make a mistake in our construction?
Fortunately not; the transducer has just not finished yet: With the input
\sageverb|14.digits(base=2)|, which is the binary expansion of~$14$ from the least
significant to the most significant digit, we would
stop in the nonfinal state with label~$2$, as we can see by typing
\begin{sageblock}
  NAF.process(14.digits(base=2))
\end{sageblock}
which results in \skipnb\sageverb|NAF.process(14.digits(base=2))|. Here, the first
component indicates whether the input is accepted or not;
the second component is the
label of the state where we stopped.
The third component of this triple is the output of the transducer
irrespective of whether the state is final or not.

By adding enough trailing zeros to the expansion of $14$, we reach a final
state and we ensure that all carries are processed. We type
\begin{sageblock}
  NAF_of_14 = NAF(14.digits(base=2) + [0, 0, 0])
\end{sageblock}
and get the output \sageverb|NAF_of_14|.
This list corresponds to the digits of the non-adjacent form
of~$14$, starting with the digit corresponding to $1$ at the
left, and then continuing with the digits corresponding to  $2$, $4$, $8$, $16$,
and $32$.

But do we really want to think about trailing zeros? This should
be done by \SageMath{}. And that is possible by
\begin{sageblock}
  NAF = NAF.with_final_word_out(0)
\end{sageblock}
This function constructs a final output for every state. The final output of
a state is appended to the ``normal'' output if we stop reading the input in
this state. Transducers with final output are called \emph{subsequential},
cf.~\cite{Schuetzenberger:1977}. The method \verb|with_final_word_out()|
computes the final output by reading as many zeros as necessary to
reach a final state (if possible). The corresponding output is then the final
output.

Now, we compute the non-adjacent form of $14$ again by
\begin{sageblock}
  NAF_of_14 = NAF(14.digits(base=2))
\end{sageblock}
without thinking about how many trailing zeros we have to add. And the result
\sageverb|NAF_of_14| is still the same as before (except for one trailing zero).


\subsection{Calculating the Non-Adjacent Form with Less Thinking}
\label{sec:NAF-transition-function}


A different approach to construct a transducer calculating the non-adjacent form
is via a transition function.

To get this function, we think about the following algorithm rewriting the
binary expansion to the
NAF: We start by determining the least significant digit~$n_{0}$ of the
non-adjacent form of the integer~$n$. This can be decided by looking at the two
least significant digits of the binary expansion: If $n$ is even, then the
digit of the non-adjacent form is zero. If $n$ is odd, it is $1$ or $\bar1$,
depending on $n$ modulo~$4$. As the next step of this algorithm, we have to
compute the non-adjacent form of $\frac12(n-n_{0})$.

We can reformulate this as a transition function.
As we have to look at two
consecutive input digits, we simply read and store the very first input digit
by inserting an additional rule for the initial state. This
leads to the following code:
\begin{sageblock}
  def NAF_transition(state_from, read):
      if state_from == 'I':
          write = None
          state_to = read
          return (state_to, write)
      current = 2*read + state_from
      if current % 2 == 0:
          write = 0
      elif current % 4 == 1:
          write = 1
      else:
          write = -1
      state_to = (current - write) / 2
      return (state_to, write)
\end{sageblock}
Here, \verb|%| is the
remainder of the integer division in \SageMath{}.

The transducer defined by this transition function can be built by
\begin{sageblock}
  NAF2 = Transducer(NAF_transition,
                    initial_states=['I'],
                    final_states=[0],
                    input_alphabet=[0, 1]).with_final_word_out(0)
\end{sageblock}
We can check whether the two transducers are the same by
\begin{sageblock}
  NAF == NAF2
\end{sageblock}
which, luckily, yields \skipnb\sageverb|NAF==NAF2|.

If we think again about our algorithm at the beginning of the section,
we see that we can also use a different input alphabet, e.g.\ \verb|[-1, 0, 1]|. Then we
obtain a transducer which can transform any binary digit expansion
with digits $-1$, $0$ and $1$ into a non-adjacent form. The code is
\begin{sageblock}
  NAF_all = Transducer(NAF_transition,
                       initial_states=['I'],
                       final_states=[0],
                       input_alphabet=[-1, 0, 1]).with_final_word_out(0)
\end{sageblock}
and you can see a picture of this transducer in
Figure~\ref{fig:naf-from-any-expansion}. Comparing \verb|NAF| in
Figure~\ref{fig:ex-transducer} and \verb|NAF_all| in
Figure~\ref{fig:naf-from-any-expansion} suggests that \verb|NAF| is
contained in \verb|NAF_all| as a subgraph because such an inclusion
holds for their digit sets as well: $\{0,1\}\subseteq\{-1,0,1\}$.

\begin{figure}
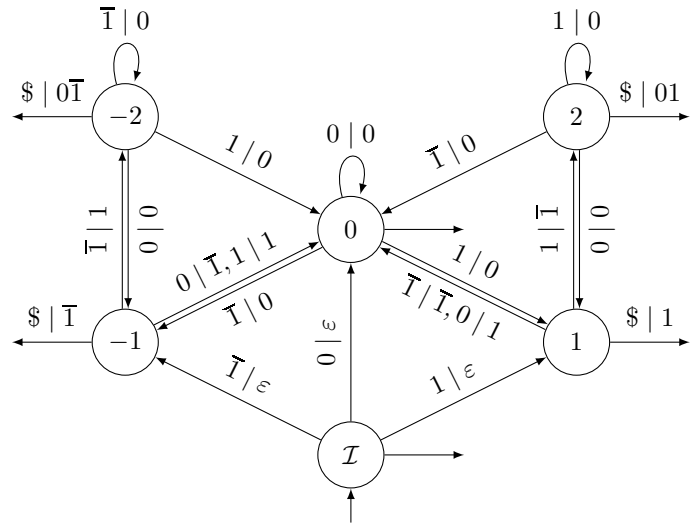

  \centering
  \begin{sagesilent}
    # Figure: A transducer transforming any binary digit expansion
    #         with digits -1, 0 and 1 into a non-adjacent form
    NAF_all.state('I').format_label=lambda: r'\mathcal{I}'
    NAF_all.latex_options(
        coordinates={'I': (0, -1.5), 0: (0, 1.5), 1: (3, 0),
                     -1: (-3, 0), 2: (3, 3), -2: (-3, 3)},
        accepting_where={-1: 'left', -2: 'left', 0: 'right'},
        initial_where={'I': 'below'},
        format_letter=NAF_all.format_letter_negative)
  \end{sagesilent}
  \sagestr{latex(NAF_all)}
  \caption{A transducer transforming any binary digit expansion with digits $-1$, $0$ and $1$ into a non-adjacent form.}
  \label{fig:naf-from-any-expansion}
\end{figure}


\subsection{Back to Automata}
\label{sec:accept-naf-from-transducer}

Now we can live up to our promise from Section~\ref{sec:non-adj-form}: We can use the
output of the transducer \verb|NAF_all| to construct the automaton
accepting non-adjacent forms once more.\footnote{We choose \texttt{NAF\_all} rather than \texttt{NAF2} (or any of the other transducers), because \texttt{NAF2}
only allows positive numbers as input.}

Now, we simply ``forget'' the input of
every transition and only consider the output labels, which we can do by
\begin{sageblock}
  NAF_language4 = NAF_all.output_projection().minimization()
\end{sageblock}
Again we can check the equivalence by
\verb|NAF_language4.is_equivalent(NAF_language)| which yields
\sageverb|NAF_language4.is_equivalent(NAF_language)|.

So far so good; we have played around with automata and transducers a
lot. But now it is time to get in touch with some other mathematical
areas. In the following section, we will consider a
minimization problem.


\subsection{What Makes the Non-Adjacent Form Better than Other Expansions?}
\label{sec:what-makes-naf-better}

To answer this question, first take a look at
Figure~\ref{fig:naf-from-any-expansion}. Forget everything you have
read up to now for a moment and use some imagination.  Does the
object in this figure remind you of anything?---A spiderweb?---Well,
let us rephrase the question: Does the object in this figure remind
you of any well-known mathematical object?---Go one or two steps
back.---Better now?---Do you see a graph? More precisely a directed
graph with labels?---Very well.

Since our mind could do this switching to a different world, \SageMath{} should be able to do this as well. We type
\begin{sageblock}
  def weight(word):
      return sum(ZZ(d != 0) for d in word)
  G = NAF_all.digraph(
      edge_labels=lambda t: weight(t.word_in) - weight(t.word_out))
\end{sageblock}
and voil\`a, here is a graph~\verb|G|. In the code above, we use the convention
that the integer values of \verb|True| and \verb|False| are $1$ and
$0$ respectively. The edge labels of \verb|G| will be explained
in a moment.

Now back to the question posed in the title of this part. Let us first
ask what ``better'' means. The answer can be stated as follows: A
\emph{minimal digit expansion} has the lowest possible number of nonzero
digits (aka the \emph{Hamming weight}) among all possible expansions of the
same number with base~$2$ and digits~$-1$, $0$ and
$1$. Now, the question is: Is the non-adjacent form a minimal digit
expansion? And this is exactly the point where graphs come into play.

We have chosen the labels of the directed graph~\verb|G| such that
they are the difference between the weights of the input and the
output of the transition.  Assume for a moment that there is a path in
the graph~\verb|G| from the initial state~$\mathcal I$ to some final
state with a negative sum of labels. In terms of the
transducer~\verb|NAF_all| this means that we have read some input with
lower number of nonzeros than the corresponding output (which is a
non-adjacent form). Therefore, the non-adjacent form is not a minimal
digit expansion; we have found an expansion (the input) with lower weight.

So much for this thought
experiment; we start a search for negative paths,
i.e, look at the shortest paths from $\mathcal I$ to any state. If all
these paths have a nonnegative weight, then the non-adjacent form is
always minimal. Such a search can be done by the Bellman--Ford
algorithm (cf.\ for example \cite{Cook-Cunningham-Pulleyblank-Schrijver:1998:combin}), which is used by \SageMath{} when we call
\begin{sageblock}
  paths = G.shortest_path_lengths('I', by_weight=True)
\end{sageblock}
\skipnb\begin{sagesilent}
  minus2 = -2
  dictionary = str(paths).replace('{','\{').replace('}','\}')
\end{sagesilent}
It returns
\begin{sageverbenv}
  \verb+  +\skipnb\sageverb|dictionary|\nbcode{paths}\nbcode{paths[-2]}
\end{sageverbenv}

So, for example, the shortest path from state $\mathcal I$ to state
$-2$ has weight \skipnb\sageverb|paths[minus2]| in the graph. This means that an
input of the transducer~\verb|NAF_all| (taking exactly this path
from~$\mathcal I$ to~$-2$) has a weight one larger than its output.

All these weights are nonnegative. So we can be relieved now because we
know that the non-adjacent form is always a good choice.


\subsection{A Third Construction of the Same Transducer}
\label{sec:NAF-minus-construction}


We are now back in our construction half-marathon.
The non-adjacent form can also be generated in the following way. We start
with the binary expansions of $\frac{3n}{2}$ and of $\frac{n}{2}$. We subtract
each digit of $\frac n2$ from the corresponding digit of $\frac{3n}2$. This
leads to a digit expansion of $n$ with digits $\{-1,0,1\}$ in base $2$. One can
prove that this digit expansion is the non-adjacent form of $n$ (cf.\
\cite{Chang-Tsao-Wu:1968}, see also
\cite[Theorem~10.2.4]{VanLint:1982}).

Note that in the following, we will add a summand
$0\cdot\frac12$ to expansions like~\eqref{eq:NAF14}. This means that we write
\begin{equation*}
  14 = (100\bar{1}0\centerdot 0)_2.
\end{equation*}
It will turn out that this is convenient, since we are working with halves at
lot in our example in the following sections.

For this construction we need a few simple transducers (as, for example one for
multiplying by~$3$ and one for performing subtraction), which we combine later
appropriately. We will also reuse these machines in a later example for the
$\frac 32$--$\frac 12$-non-adjacent form in Section~\ref{sec:32-12-NAF}.

So let us start with the times-$3$-transducer, i.e., one that takes the binary expansion
of a number~$n$ as input and outputs~$3n$ (in binary). We do this, as above, by a
transition function. We define
\begin{sageblock}
  def f(state_from, read):
      current = 3*read + state_from
      write = current % 2
      state_to = (current - write) / 2
      return (state_to, write)
\end{sageblock}
to compute the next output digit (\texttt{write}) and the new carry (encoded in
\texttt{state\_to}) from the input digit (\texttt{read}) and the previous carry
(\texttt{state\_from}) in the multiplication-by-$3$-algorithm. From this
transition function we get the following transducer:
\begin{sageblock}
  Triple = Transducer(f, input_alphabet=[0, 1],
                      initial_states=[0],
                      final_states=[0]).with_final_word_out(0)
\end{sageblock}
Eager as we are, we test this construction by
\begin{sageblock}
  three_times_fourteen = Triple(14.digits(base=2))
\end{sageblock}
and get \sageverb|three_times_fourteen|, which equals~$42$. Hooray!

Back to business; our goal is to calculate binary-$3n$ minus binary-$n$. To do
so, we need a transducer which acts as identity (for the binary-$n$-part),
i.e., writes out everything that is read in. Here,
\begin{sageblock}
  Id = Transducer([(0, 0, 0, 0), (0, 0, 1, 1)],
                  initial_states=[0], final_states=[0],
                  input_alphabet=[0, 1])
\end{sageblock}
does the trick. We also get the above
by
\begin{sageblock}
  prebuiltId = transducers.Identity(input_alphabet=[0, 1])
\end{sageblock}
where we just have to specify the alphabet \verb|[0, 1]|.

As a next step (before we are heading to subtraction), we want a transducer which
produces pairs of the digits of $3n$ and of $n$ simultaneously. This can be
achieved with
\begin{sageblock}
  Combined_3n_n = Triple.cartesian_product(Id).relabeled()
\end{sageblock}

Let us test
this machine by
\begin{sageblock}
  fortytwo_and_fourteen = Combined_3n_n(14.digits(base=2))
\end{sageblock}
It returns \sageverb|fortytwo_and_fourteen|, which seems to be correct.

We further construct a transducer computing the component-wise difference: Its
input is a pair like the output of \verb|Combined_3n_n| and the output is the
difference of the two entries. We generate the transducer by
\begin{sageblock}
  Minus = transducers.operator(
      lambda read_3n, read_n: ZZ(read_3n) - ZZ(read_n),
      input_alphabet=[None, -1, 0, 1])
\end{sageblock}
where the \verb|lambda|-function specifies our operator.
Here we use that \verb|ZZ(None)| is \sageverb|ZZ(None)|.

Of course, there is not only a prebuilt identity transducer, but also a
prebuilt transducer for component-wise difference, available as
\begin{sageblock}
  prebuiltMinus = transducers.sub([-1, 0, 1])
\end{sageblock}
But unfortunately, it can only work with numbers, and we also want to
subtract \verb|None|. The final outputs are the reason: Sometimes, one
component is \verb|None|. For example, the final
output of state \sageverb|Combined_3n_n.state(1)| is
\begin{sageblock}
  final_word_out = Combined_3n_n.state(1).final_word_out
\end{sageblock}
which yields \sageverb|final_word_out|.

Finally, by
\begin{sageblock}
  NAF3 = Minus(Combined_3n_n).relabeled()
\end{sageblock}
we obtain a transducer computing the non-adjacent form of $3n-n=2n$. This
means, \verb|NAF3| is built as the composition of \verb|Minus| and
\verb|Combined_3n_n|, which could also have been called by using the method
\verb|composition()|. 

Let us test this construction. For example,
\begin{sageblock}
  NAF_of_14 = NAF3(14.digits(base=2))
\end{sageblock}
returns \sageverb|NAF_of_14|.
This is, once again, the non-adjacent form expansion of $14$,
see~\eqref{eq:NAF14}, but now starting with the digit corresponding to $\frac
12$ (which is obviously $0$) at the left, and then continuing with the digits
corresponding to $1$, $2$, $4$, $8$ and $16$.

Now we have finished our warm-up and are ready for the main example, which
will be dealing with $\frac32$--$\frac12$-non-adjacent form.


\section{An Example: Three-Half--One-Half-Non-Adjacent Form}
\label{sec:32-12-NAF}


\skipnb\subsection{What Is the Three-Half--One-Half-Non-Adjacent Form?}
\label{sec:desc-32-12-NAF}


We have (or, at least, we can calculate) the non-adjacent forms of
$\frac{3n}{2}$ and of $\frac{n}{2}$. Inspired by the construction of the NAF
presented in Section~\ref{sec:NAF-minus-construction} (as a remainder: we
subtracted two binary expansions), we build the difference of these two NAFs
as well. Thus,
we define the \emph{$\frac32$--$\frac12$-non-adjacent form} of an integer $n$
as the digit expansion obtained by subtracting each digit of the NAF of
$\frac{n}{2}$ from the corresponding digit of the NAF of $\frac{3n}{2}$. This
leads to a digit expansion of $n$ with digits $\{-2,-1,0,1,2\}$ in base
$2$. For example, the $\frac32$--$\frac 12$-non-adjacent form of---guess which
number comes now---$14$ is
\begin{equation*}
  14 = (10101\centerdot 0)_2 - (100\bar{1}\centerdot 0)_2
  = (1\bar{1}102\centerdot 0)_2.
\end{equation*}

In a first step, we want to calculate this new expansion; see the following
sections. On the one hand, we are lazy and want to reuse as much as possible
from the finite state machines we have already constructed. On the other hand, we are
motivated to use our new knowledge on working with those automata and
transducers. So the idea will be to combine several of these known transducers
appropriately.


\subsection{Combining Small Transducers to a Larger One}
\label{sec:construction-32-12-NAF}


We first combine the transducers \verb|Triple| and \verb|NAF| to obtain a
transducer to compute the non-adjacent form of $3n$. For convenience, we
choose
 \verb|NAF = NAF3|, because there we do not have to consider an empty
output of a transition.
\begin{sagesilent}
  NAF = NAF3
\end{sagesilent}

We use
\begin{sageblock}
  NAF3n = NAF(Triple)
\end{sageblock}
which builds the composition of the two transducers involved and therefore
gives us a gadget to get the non-adjacent form of $3n$.

Next, we construct a transducer which builds the non-adjacent forms of $3n$
and $n$ simultaneously by
\begin{sageblock}
  Combined_NAF_3n_n = NAF3n.cartesian_product(NAF).relabeled()
\end{sageblock}
The function \texttt{cartesian\_product()} sounds familiar---we used it in
Section~\ref{sec:NAF-minus-construction} already. It constructs a transducer
which writes pairs of digits.

Finally, by reusing \verb|Minus|, we construct
\begin{sageblock}
  T = Minus(Combined_NAF_3n_n).relabeled()
\end{sageblock}
This transducer finally computes the $\frac32$--$\frac12$-non-adjacent form.
To get some information like the number of states of the finite state machine,
we type \verb|T| in \SageMath{} and see
\begin{sageverbenv}
  \verb+  +\sageverb|T|
\end{sageverbenv}

Let us continue with the example from the beginning. To compute this new digit
expansion of $14$, we type
\begin{sageblock}
  expansion_of_14 = T(14.digits(base=2))
\end{sageblock}
The output is
\begin{sageverbenv}
  \verb+  +\sageverb|expansion_of_14|
\end{sageverbenv}
which is the $\frac 32$--$\frac 12$-non-adjacent form of $14$ starting with
the digit corresponding to $\frac 14$. This starting digit has the following reasons:
The output of the transducer \verb|NAF| starts with the digit corresponding
to $\frac 12$ when reading $n$. We use the non-adjacent form of $\frac n2$,
which thus starts at the digit corresponding to $\frac 14$.


\subsection{Getting a Picture}
\label{sec:plot}


Up to now, we constructed a couple of (or one might call it ``many'')
finite state machines, but we barely saw one. Okay, to be fair, we
have drawn an automaton in Section~\ref{sec:recogn-naf}.---But how did
this work exactly?

With \verb|T.plot()| we get a
first graphical representation of the transducer. This was easy, but we are not
fully satisfied. For example, the labels of the transitions are missing. And,
maybe we want to rearrange the states a little bit to obtain less crossings of
the transitions. This can be achieved by the following: 
\verb|view(T)| gives a second graphical representation of the
transducer. Maybe the arrangement of the states is not nicer than before, but we will improve this a lot.

We first choose the coordinates of the states by
\begin{sageblock}
  T.set_coordinates({
      0: (-2, 0.75),
      1: (0, -1),
      2: (-6, -1),
      3: (6, -1),
      4: (-4, 2.5),
      5: (-6, 5),
      6: (6, 5),
      7: (4, 2.5),
      8: (2, 0.75)})
\end{sageblock}
Furthermore, for transition labels, we prefer ``$\overline{1}$'' over ``$-1$'',
so we choose the appropriate formatting function. Additionally, we choose the
directions of the arrows with the final outputs.
\begin{sageblock}
  T.latex_options(format_letter=T.format_letter_negative,
                  accepting_where={
                    0: 'right',
                    1: 'below',
                    2: 'below',
                    3: 'below',
                    4: 60,
                    5: 'above',
                    6: 'above',
                    7: 120,
                    8: 'left'},
                  accepting_show_empty=True)
\end{sageblock}
Now, the output of \verb|view(T)| in \SageMath{} looks like
Figure~\ref{fig:T}. The $\$ $-symbol signals the end of the input sequence.
Further customization of the underlying \TikZ{}-code is possible, see the documentation of
\verb|latex_options()|.

On the other hand, by typing \verb|latex(T)| we get this \TikZ{}-code for the
transducer, which can be used to include a figure of the finite state machine
in a \LaTeX-document, like it was done for this tutorial. To succeed, we need to
use the package \texttt{tikz} and to include the line \verb|\usetikzlibrary{automata}| in
the preamble of the \LaTeX-document.

\begin{figure}
  \centering
  \sagestr{latex(T)}
  \caption{Transducer \texttt{T} to compute the
    $\frac32$--$\frac12$-non-adjacent form of $n$.}
  \label{fig:T}
\end{figure}


\subsection{Accepting More Strange Animals}
\label{sec:recognizing}


As for the non-adjacent form
(Sections~\ref{sec:accept-naf-from-transducer}),
we can construct an automaton accepting all 
$\frac32$--$\frac12$-non-adjacent forms by
\begin{sageblock}
  R = T.output_projection().minimization()
\end{sageblock}
It has \sageverb|len(R.states())| states, so we do not show it here. However, we
can still compute the number of $\frac32$--$\frac12$-non-adjacent forms of
length $n$ (see also Section~\ref{sec:counting}). A simple
\begin{sageblock}
  N = R.number_of_words(n, CyclotomicField(3))
\end{sageblock}
results in
\begin{equation*}
  \sage{N}.
\end{equation*}
What was that \verb+CyclotomicField(3)+ about? It was a hint for \SageMath{} to
compute in that number field instead of computing in the field of algebraic
numbers. This is more efficient and produces nicer output.

Let us persuade ourselves that the answer is correct by considering
the first few values.
\begin{sageblock}
  bool(sum(N.subs(n=i) for i in srange(6)) == len(list(R.language(5))))
\end{sageblock}
\skipnb\begin{sagesilent}
  test = bool(sum(N.subs(n=i) for i in srange(6)) ==
              len(list(R.language(5))))
\end{sagesilent}
yields \skipnb\sageverb|test|. In fact,
\verb+R.language(5)+ iterates over all words accepted by \verb+R+ of length
at most $5$.


\subsection{(Heavy) Weights}
\label{sec:weights}


Our original motivation to study the $\frac32$--$\frac12$-non-adjacent form
comes from analyzing its (Hamming) weight, i.e., the number of nonzero
digits. We want to compare the Hamming weights of the different digit
expansions: standard binary expansion, non-adjacent form and $\frac
32$--$\frac 12$-non-adjacent form.

Using the ideas of the constructions above, this is also not difficult. We
construct a transducer computing the weight of the input by
\begin{sageblock}
  def weight(state_from, read):
      write = ZZ(read != 0)
      return (0, write)
  Weight = Transducer(weight, input_alphabet=srange(-2, 2+1),
                      initial_states=[0], final_states=[0])
\end{sageblock}
The transducer \verb|Weight|
writes a $1$ for every nonzero input, which means that the weight is encoded
in unary in the output string.

There also exists a prebuilt transducer which we could use instead of our own
construction. It is available via
\begin{sageblock}
  prebuiltWeight = transducers.weight(srange(-2, 2+1))
\end{sageblock}

Composing the weight-transducer with the one calculating the
$\frac32$--$\frac12$-NAF by
\begin{sageblock}
  W = Weight(T)
\end{sageblock}
we end up with a transducer with \sageverb|len(W.states())|~states computing the
Hamming weight of this new digit expansion of~$n$ (given in binary). For
instance,
\begin{sageblock}
  W(14.digits(base=2))
\end{sageblock}
yields \skipnb\sageverb|W(14.digits(base=2))|, which means the weight is
\sageverb|add(W(14.digits(base=2)))|.

The transducer \verb|W| can be further simplified by
\begin{sageblock}
  W = W.simplification()
\end{sageblock}
Why did we not use \verb|minimization()| as we did in
Section~\ref{sec:non-adj-form}? The reason is that we work with
transducers now, in contrast to Section~\ref{sec:non-adj-form}, where
we used automata. There is not always a unique minimal
transducer for a given transducer
(cf.~\cite{Choffrut:2003:minim}). But we can at least simplify the
transducer to obtain one with a
smaller number of states.

If you wonder, why there is the word ``heavy'' in the title of this part, read
on until the end of the example.


\subsection{Also Possible: Adjacency Matrices}
\label{sec:adj-matrix}


We want to asymptotically analyze the expected value of the Hamming weight of our new digit
expansion for all positive integers less than $2^{k}$, where $k$ is a fixed
large number.

We dive a bit deeper into the mathematics involved now.
One way to perform the asymptotic analysis is by means of the
adjacency matrix of the transducer. By
\begin{sageblock}
  var('y')
  def am_entry(trans):
      return y^add(trans.word_out) / 2
  A = W.adjacency_matrix(entry=am_entry)
\end{sageblock}
we obtain a matrix where the entry at $(k,l)$ is $\frac 12 y^{h}$ if there is a
transition with output sum $h$ from state $k$ to $l$ and $0$ otherwise. The
adjacency matrix generated by this command is
\begin{sagesilent}
    latex.matrix_column_alignment('c')
\end{sagesilent}
\begin{equation*}
  A = \sage{A}.
\end{equation*}

For $y=1$, this is simply the transition probability matrix. Its normalized
left eigenvector to the eigenvalue $1$ gives the stationary
distribution. We write
\begin{sageblock}
  dim = len(W.states())
  I = matrix.identity(dim)
  (v_not_normalized,) = (A.subs(y=1) - I).left_kernel().basis()
  v = v_not_normalized / v_not_normalized.norm(p=1)
\end{sageblock}
and obtain \sageverb|v|.

To obtain the average Hamming weight of the  $\frac 32$--$\frac
12$-non-adjacent form, we compute the expected output vector in each state
as
\begin{sageblock}
  expected_output = derivative(A, y).subs(y=1) * vector(dim*[1])
\end{sageblock}
and obtain $\sage{expected_output}$.
Note that the derivative here simply computes the expected output for every
transition. We could also have called \verb|adjacency_matrix()| with a suitably
modified entry function.

The expected density is therefore
\skipnb\begin{sageblock}
  v * expected_output
\end{sageblock}
which yields $\sage{v * expected_output}$. This means that the main term of the
average number of nonzero digits in $\frac 32$--$\frac 12$-NAFs of length~$k$
is $\skipnb\sage{v * expected_output} k$.


\subsection{More on the Hamming Weight by Letting SageMath Do the Work}
\label{sec:weight-more}

We have the impression that the analysis of the previous section could be done (or, rephrased, we
want that this should be done) more automatically. Indeed, we can let \SageMath{}
do the work for us, and it does it very well: It not only outputs the mean of
the Hamming weight, but also its variance and more.

By
\begin{sageblock}
  var('k')
  moments = W.asymptotic_moments(k)
\end{sageblock}
we obtain a dictionary whose entries are the expectation and the variance of the
sum of the output of the transducer, and the covariance of the sum of the
output and the input of the transducer (cf.\ 
\cite{Heuberger-Kropf-Wagner:2014:combin-charac}). The probability model is the
equidistribution on all input sequences of a fixed length $k$.

The expected value of the Hamming weight
of the $\frac 32$--$\frac 12$-non-adjacent form is
\begin{equation*}
  \sage{moments['expectation']}.
\end{equation*}
as $k$ tends to infinity.

This function can also give us the variance of the Hamming weight of the
$\frac 32$--$\frac 12$-non-adjacent form, which is
\begin{equation*}
  \sage{moments['variance']}.
\end{equation*}

Of course we could do a lot more beautiful stuff. We could construct a
bivariate generating function. From this, we could obtain more terms and better error terms of the
asymptotic expansion of the expected value, the variance and higher
moments. We could also prove a central limit theorem. And everything by using
the full power of \SageMath{}. But, again, we do not want
to go into details here. 
We refer to the book by Flajolet and
Sedgewick~\cite{Flajolet-Sedgewick:ta:analy} for details on the asymptotic
analysis of digit expansions and other sequences.


\subsection{What Does This Mean for This Brand New Digit Expansion?}
\label{sec:conclusion}


In the past several sections, we were able to calculate the average
Hamming weight of the $\frac32$--$\frac12$-non-adjacent form asymptotically by
the help of the finite state machines package in \SageMath{}. But what does this
result tell us?

So let us compare this digit expansion with the standard binary expansion and
the classical non-adjacent form. The expected value of the Hamming weight of
the standard binary expansion can be calculated by
\begin{sageblock}
  expectation_binary = Id.asymptotic_moments(k)['expectation']
\end{sageblock}
which gives
\begin{equation*}
  \sage{expectation_binary}.
\end{equation*}
Of course, it was not necessary to use the transducer \verb|Weight| here (since
we only have digits $0$ and $1$). The expected value of the weight of the NAF
can be obtained with the code
\begin{sageblock}
  expectation_NAF = Weight(NAF).asymptotic_moments(k)['expectation']
\end{sageblock}
which produces the weight
\begin{equation*}
  \sage{expectation_NAF},
\end{equation*}
cf.\ also~\cite{Morain-Olivos:1990}. Note that in this particular construction
(only digits $-1$, $0$ and $1$), we could have used the prebuilt transducer
\begin{sageblock}
  Abs = transducers.abs([-1, 0, 1])
\end{sageblock}
instead of the weight-transducer.

Both values are (asymptotically) less than the
\begin{equation*}
  \sage{moments['expectation']}
\end{equation*}
of the $\frac32$--$\frac12$-non-adjacent form, which means that this expansion
has much more nonzero digits on average and therefore is much ``heavier''. So, from a point of view of minimizing the Hamming weight, this new
expansion is disappointing: It uses more digits but realizes a larger weight.


\skipnb\section{More Questions and Answers}


\skipnb\subsection{Does This Automata-Package Have a History?}


Indeed it has. One of the authors wrote a similar
unpublished package for Mathematica~\cite{Mathematica:2005},
which is used in the articles
\cite{%
  Avanzi-Heuberger-Prodinger:2006:scalar-multip-koblit-curves,
  Avanzi-Heuberger-Prodinger:2010:arith-of,
  Avanzi-Heuberger-Prodinger:2011:redun-expan-i,
  Grabner-Heuberger-Prodinger:2005:counting-optimal-joint,
  Heuberger:2004:minim-expan-fibonacci-greedy,
  Heuberger:2010:nonoptimality,
  Heuberger-Katti-Prodinger-Ruan:2005:altgreedy,
  Heuberger-Muir:2007:minim-weigh,
  Heuberger-Prodinger:2006:analy-alter,
  Heuberger-Prodinger:2007:hammin-weigh,
  Heuberger-Prodinger:2007:expan-number,
  Heuberger-Prodinger:2009:analy-compl,
  Heuberger-Prodinger-Wagner:2008:posit-number}.
But this implementation was of limited
scope and the rise of \SageMath{} made it clear that it is ripe for a redesigned
version. Having the finite state machine module readily available in a publicly
available and continuously maintained system also leads to more transparency in
the computational parts of publications.


\skipnb\subsection{What about Other Finite State Machine Packages?}


There are quite a lot of implementations for finite state machines available.
One of the fastest libraries is OpenFST~\cite{OpenFst:2013}, for which a Python
interface~\cite{pyopenfst:2013} exists, too. But since it is written in C/C++,
it does not work well with the mathematical objects defined in \SageMath{}.
Vcsn (Vaucanson)~\cite{2015:vcsn} is written in C++ and offers a
Python interface as well.
Other
non-Python modules are, for example, \cite{ASTL:2013, brics:2011,
  SFST:2013}. There also exist a couple of Python packages,
e.g.~\cite{automata:2013, fsa:2004, python-automata:2007}, which can also be
found on the Python-Wiki.\footnote{The Python-Wiki can be found at
  \url{https://wiki.python.org/moin/FiniteStateMachine}.}  Some of those are
specialized (and thus not flexible enough) and implement only partial support
for both automata and transducers. It seems that some of them are even
out-dated and not developed any further.


\skipnb\subsection{How Fast is This Package?}
\label{sec:how-fast}


To get a feeling, let us have a look at the following concrete example:
We construct a minimal automaton
forbidding the subwords $10^n1$ for all $1\leq n<15$. We can achieve
this in the same manner as in Section~\ref{sec:recogn-naf-complement}---there
we have forbidden adjacent nonzeros---by
\begin{sageverbenv}
\begin{verbatim}
  S = [automata.ContainsWord('o' + n*'z' + 'o', input_alphabet=['o', 'z'])
       for n in range(1, 15)]
  P = sum(S[1:], S[0])
  A = P.determinisation().complement()
  M = A.coaccessible_components().minimization()
\end{verbatim}
\end{sageverbenv}
While \verb|P| has only $147$ states, the automaton~\verb|A| consists
of $245\,760$ states. Due to this large number of
states, it takes \SageMath{} roughly three minutes to create \verb|A| out
of \verb|P|. In comparison, the compiled automaton\footnote{The equivalent code
  of the example in Section~\ref{sec:how-fast} using the Python bindings of
  Vcsn~\cite{2015:vcsn} is
\begin{sageverbenv}
\begin{verbatim}
  import vcsn
  context = vcsn.context("lal_char(oz), b")
  any_word = context.expression('\z').automaton().complete().complement()
  S = [any_word * context.expression('o' + n*'z' + 'o').automaton() * any_word
       for n in range(1, 15)]
  P = sum(S[1:], S[0])
  A = P.complete().determinize().complement()
  M = A.trim().minimize()
\end{verbatim}
\end{sageverbenv}}
of the C-library
Vcsn~\cite{2015:vcsn} achieves this in slightly less than one tenth of
the time.
The resulting minimal automaton $M$ has $16$ states;
its creation out of $A$ is not time-critical.

 Of course, speed comparisons between an interpreted, dynamic
programming language such as Python and an optimized C-library can only have
one outcome. One main design goal of the \SageMath{} package is flexibility and a good interplay with mathematical
objects. In fact, all labels of states and transitions are allowed to
be arbitrary \SageMath-objects (and not only letters or strings
as in the example shown here). Another
design goal was to have a very low entry barrier for using the
package due to its inclusion in \SageMath.


\skipnb\subsection{How Complex is Everything?}


Many algorithms for automata have an
exponential worst case running time. (Think of determinisation!) However, we tried
to choose efficient algorithms whenever possible. For instance, \SageMath's
package
implements Moore's algorithm for minimization of deterministic
automata, which has a polynomial worst case complexity
\cite[Sec.~I.3.3.3]{Sakarovitch:2009:elemen}. Additionally, the user may also choose Brzozowski's
algorithm. This algorithm
has an exponential worst case complexity but performs well in
empirical experiments \cite{Almeida-Moreira-Reis:2008}, and it works
with nondeterministic automata, too.


\skipnb\subsection{What will the Future Bring Us?}


Of course, nobody knows---but there is room for improvement and
extensions. Performance of the package might be improved by 
Cythonizing\footnote{Cython (semi-)automatically translates Python
  code to C code and then compiles it.} the source code or by calling
specialized packages for time-critical operations.


\egroup

\bibliographystyle{amsplain-sort-by-keys}
\bibliography{bib/cheub}

\providecommand{\Submitted}{Submitted} \providecommand{\availableat}{ available
  at } \providecommand{\alsoavailableat}{ also available at }
  \providecommand{\evavailableat}{earlier version available at }
  \providecommand{\toappearin}{To appear in } \providecommand{\toappear}{to
  appear} \providecommand{\inpreparation}{in preparation}
  \providecommand{\doi}[1]{\href{http://dx.doi.org/#1}{\path{doi:#1}}}
  \providecommand{\etc}{\emph{etc.}}\def\cprime{$'$}
\providecommand{\bysame}{\leavevmode\hbox to3em{\hrulefill}\thinspace}
\providecommand{\MR}{\relax\ifhmode\unskip\space\fi MR }
\providecommand{\MRhref}[2]{%
  \href{http://www.ams.org/mathscinet-getitem?mr=#1}{#2}
}
\providecommand{\href}[2]{#2}
\begin{thebibliography}{10}

\bibitem{Almeida-Moreira-Reis:2008}
Marco Almeida, Nelma Moreira, and Rog{\'e}rio Reis, \emph{On the performance of
  automata minimization algorithms}, CiE 2008: Abstracts and extended abstracts
  of unpublished papers, 2008.

\bibitem{ASTL:2013}
\emph{The automata standard template library},
  \url{http://astl.sourceforge.net}, 2013.

\bibitem{automata:2013}
\emph{{automata} 0.1.4}, \url{https://pypi.python.org/pypi/automata}, 2013.

\bibitem{Avanzi-Heuberger-Prodinger:2006:scalar-multip-koblit-curves}
Roberto Avanzi, Clemens Heuberger, and Helmut Prodinger, \emph{Scalar
  multiplication on {K}oblitz curves. {U}sing the {F}robenius endomorphism and
  its combination with point halving: {E}xtensions and mathematical analysis},
  Algorithmica \textbf{46} (2006), 249--270. \MR{2291956 (2008a:94180)}

\bibitem{Avanzi-Heuberger-Prodinger:2010:arith-of}
\bysame, \emph{Arithmetic of supersingular {K}oblitz curves in characteristic
  three}, Cryptology ePrint Archive, Report 2010/436, 2010.

\bibitem{Avanzi-Heuberger-Prodinger:2011:redun-expan-i}
\bysame, \emph{Redundant $\tau$-adic expansions {I}: Non-adjacent digit sets
  and their applications to scalar multiplication}, Des. Codes Cryptogr.
  \textbf{58} (2011), 173--202. \MR{2770310 (2012f:11010)}

\bibitem{brics:2011}
\emph{{dk.brics.automaton} 1.11-8}, \url{http://www.brics.dk/automaton/}, 2011.

\bibitem{Chang-Tsao-Wu:1968}
Sze-hou Chang and Nelson~T. Tsao-Wu, \emph{Distance and structure of cyclic
  arithmetic codes}, Proc. {H}awaii International Conference on System
  Sciences, vol.~1, 1968, pp.~463--466.

\bibitem{Choffrut:2003:minim}
Christian Choffrut, \emph{Minimizing subsequential transducers: a survey},
  Theoret. Comput. Sci. \textbf{292} (2003), no.~1, 131--143, Selected Papers
  in honor of Jean Berstel.

\bibitem{Cook-Cunningham-Pulleyblank-Schrijver:1998:combin}
William~J. Cook, William~H. Cunningham, William~R. Pulleyblank, and Alexander
  Schrijver, \emph{Combinatorial optimization}, Wiley-Interscience Series in
  Discrete Mathematics and Optimization, John Wiley \& Sons Inc., New York,
  1998. \MR{99b:90098}

\bibitem{Flajolet-Sedgewick:ta:analy}
Philippe Flajolet and Robert Sedgewick, \emph{Analytic combinatorics},
  Cambridge University Press, Cambridge, 2009.

\bibitem{fsa:2004}
\emph{{FSA} -- {F}inite {S}tate {A}utomaton processing in {P}ython},
  \url{http://www.osteele.com/software/python/fsa/}, 2004.

\bibitem{Grabner-Heuberger-Prodinger:2005:counting-optimal-joint}
Peter~J. Grabner, Clemens Heuberger, and Helmut Prodinger, \emph{Counting
  optimal joint digit expansions}, Integers \textbf{5} (2005), no.~3, A9.
  \MR{2191755 (2006i:11008)}

\bibitem{Heuberger:2004:minim-expan-fibonacci-greedy}
Clemens Heuberger, \emph{Minimal expansions in redundant number systems:
  Fibonacci bases and greedy algorithms}, Period. Math. Hungar. \textbf{49}
  (2004), 65--89. \MR{2106466 (2005m:11009)}

\bibitem{Heuberger:2010:nonoptimality}
\bysame, \emph{Redundant $\tau$-adic expansions {II}: {N}on-optimality and
  chaotic behaviour}, Math. Comput. Sci. \textbf{3} (2010), 141--157.
  \MR{2608292 (2011b:11012)}

\bibitem{Heuberger-Katti-Prodinger-Ruan:2005:altgreedy}
Clemens Heuberger, Rajendra Katti, Helmut Prodinger, and Xiaoyu Ruan, \emph{The
  alternating greedy expansion and applications to left-to-right algorithms in
  cryptography}, Theoret. Comput. Sci. \textbf{341} (2005), 55--72. \MR{2159644
  (2007d:94034)}

\bibitem{Heuberger-Krenn-Kropf:2013:fsm-sage}
Clemens Heuberger, Daniel Krenn, and Sara Kropf, \emph{Finite state machines,
  automata, transducers}, \url{http://trac.sagemath.org/15078}, 2013, module in
  \href{http://www.sagemath.org/}{Sage 5.13}.

\bibitem{Heuberger-Kropf-Wagner:2014:combin-charac}
Clemens Heuberger, Sara Kropf, and Stephan Wagner, \emph{Variances and
  covariances in the central limit theorem for the output of a transducer},
  European J. Combin. \textbf{49} (2015), 167--187. \MR{3349532}

\bibitem{Heuberger-Muir:2007:minim-weigh}
Clemens Heuberger and James~A. Muir, \emph{Minimal weight and
  colexicographically minimal integer representations}, J. Math. Cryptol.
  \textbf{1} (2007), 297--328. \MR{2441062 (2010f:11014)}

\bibitem{Heuberger-Prodinger:2006:analy-alter}
Clemens Heuberger and Helmut Prodinger, \emph{Analysis of alternative digit
  sets for nonadjacent representations}, Monatsh. Math. \textbf{147} (2006),
  219--248. \MR{2215565 (2007g:11010)}

\bibitem{Heuberger-Prodinger:2007:hammin-weigh}
\bysame, \emph{The {H}amming weight of the non-adjacent-form under various
  input statistics}, Period. Math. Hungar. \textbf{55} (2007), 81--96.
  \MR{2341895 (2009b:05021)}

\bibitem{Heuberger-Prodinger:2007:expan-number}
\bysame, \emph{On $\alpha$-greedy expansions of numbers}, Adv. in Appl. Math.
  \textbf{38} (2007), 505--525. \MR{2311049 (2008b:11011)}

\bibitem{Heuberger-Prodinger:2009:analy-compl}
\bysame, \emph{Analysis of complements in multi-exponentiation algorithms using
  signed digit representations}, Internat. J. Found. Comput. Sci. \textbf{20}
  (2009), 443--453. \MR{2533269}

\bibitem{Heuberger-Prodinger-Wagner:2008:posit-number}
Clemens Heuberger, Helmut Prodinger, and Stephan~G. Wagner, \emph{Positional
  number systems with digits forming an arithmetic progression}, Monatsh. Math.
  \textbf{155} (2008), 349--375. \MR{2461584 (2009j:11018)}

\bibitem{Hopcroft-Motwani-Ullman:2001:automata}
John~E. Hopcroft, Rajeev Motwani, and Jeffrey~D. Ullman, \emph{Introduction to
  automata theory, languages, and computation}, Addison-Wesley series in
  computer science, Addison-Wesley, 2001.

\bibitem{Morain-Olivos:1990}
Fran\c{c}ois Morain and Jorge Olivos, \emph{Speeding up the computations on an
  elliptic curve using addition-subtraction chains}, RAIRO Inform. Th\'eor.
  Appl. \textbf{24} (1990), 531--543.

\bibitem{OpenFst:2013}
\emph{{OpenFst} {L}ibrary 1.5.0}, \url{http://openfst.org}, 2015.

\bibitem{pyopenfst:2013}
\emph{{pyopenfst}}, \url{https://github.com/tmbdev/pyopenfst}, 2014.

\bibitem{python-automata:2007}
\emph{{python-automata} 1.0}, \url{https://code.google.com/p/python-automata/},
  2007.

\bibitem{Reitwiesner:1960}
George~W. Reitwiesner, \emph{Binary arithmetic}, Advances in Computers, {V}ol.
  1, Academic Press, New York, 1960, pp.~231--308. \MR{0122018 (22 \#12745)}

\bibitem{SageMath:2015:6.10}
The~SageMath Developers, \emph{{SageMath} {M}athematics {S}oftware ({V}ersion
  6.10)}, 2015, \url{http://www.sagemath.org}.

\bibitem{Stein-others:2013:sage-mathem-5.13}
William~A. Stein et~al., \emph{{S}age {M}athematics {S}oftware ({V}ersion
  5.13)}, The Sage Development Team, 2013, \url{http://www.sagemath.org}.

\bibitem{Sakarovitch:2009:elemen}
Jacques Sakarovitch, \emph{Elements of automata theory}, Cambridge University
  Press, Cambridge, 2009, Translated from the 2003 French original by Reuben
  Thomas.

\bibitem{Schuetzenberger:1977}
Marcel-Paul Sch\"utzenberger, \emph{Sur une variante des fonctions
  sequentielles}, Theoret. Comput. Sci. \textbf{4} (1977), no.~1, 47--57.

\bibitem{SFST:2013}
\emph{{SFST} 1.4.7a}, \url{http://www.cis.uni-muenchen.de/~schmid/tools/SFST}/,
  2015.

\bibitem{VanLint:1982}
Jacobus~Hendricus van Lint, \emph{Introduction to coding theory}, Graduate
  Texts in Mathematics, vol.~86, Springer, 1992.

\bibitem{2015:vcsn}
\emph{Vcsn 2.1}, \url{https://www.lrde.epita.fr/wiki/Vcsn}, 2015.

\bibitem{Mathematica:2005}
{Wolfram Research, Inc.}, \emph{{M}athematica ({V}ersion 5.2)}, 2005.

\end{thebibliography}

\end{document}